\documentstyle[aps,pre,epsfig]{revtex}\twocolumn
\newcommand{\T}{{\cal T}}
\newcommand{\Ord}{{\cal O}}

\newcommand{\PA}{P_{\cal A}}

\newcommand{\J}{\Lambda}

\begin{document}
\title{Surmounting Oscillating Barriers}
\author{J\"org Lehmann, Peter Reimann,  and Peter H\"anggi}
\address{Universit\"at Augsburg, 
         Institut f\"ur Physik,
         Universit\"atsstr.~1, 
         D-86135 Augsburg, 
         Germany}
\maketitle
\begin{abstract}
  Thermally activated escape over a potential barrier in the presence of
  periodic driving is considered.  By means of novel time-dependent
  path-integral methods we derive asymptotically exact weak-noise expressions
  for both the {\em instantaneous} and the {\em time-averaged} escape rate.  
  The agreement
  with accurate numerical results is excellent over a wide range of 
  driving strengths and driving frequencies.
\end{abstract}
\vspace{2mm}

The problem of noise driven escape over a potential barrier
is ubiquitous in natural sciences \cite{han90}. 
Typically, the noise is weak and the escape time is governed by 
an exponentially leading Arrhenius factor. 
This scheme, however, meets formidable difficulties in far from equilibrium 
systems due to the extremely complicated interplay between 
{\em global} properties of the metastable potential and the noise
\cite{han90,noneq}. Prominent examples are systems 
driven by time-periodic forces \cite{limits,sme99},
exemplified by strong laser driven semiconductor
heterostructures, stochastic resonance \cite{sr}, directed
transport in rocked Brownian motors \cite{rat}, or
periodically driven ``resonant activation'' processes \cite{ra}
like AC driven biochemical reactions in protein membranes.  
Despite its experimental importance, the theory of
oscillating barrier crossing is still in its infancy. 
Previous attempts have been restricted to
weak (linear response), slow (adiabatic regime),
or fast (sudden regime) driving \cite{limits,sme99,sr}.
In this Letter we address the most challenging regime
of {\it strong and moderately fast} driving
by means of path-integral methods. In fact, 
our approach becomes asymptotically exact 
for any finite amplitude and period of the driving 
as the noise strength tends to zero, and comprises a conceptionally 
new, systematic treatment of the {\it rate prefactor} multiplying 
the exponentially leading Arrhenius factor.
Closest in spirit is the
recent work  \cite{sme99}, which is restricted, however, 
to the linear response regime for the exponentially leading part and treats
the prefactor with a matching procedure, involving the barrier region
only.
Our analytical theory is tested for a sinusoidally rocked
metastable potential against very
precise numerical results.
Conceptionally, our approach should be of considerable
interest for many related problems:  
generalizations for higher dimensional systems and for
non-periodic driving forces.

%
%

{\em Model ---}We consider the overdamped escape dynamics of a
Brownian particle $x(t)$ in properly scaled units
\begin{equation}
\dot x(t) = F(x(t),t)+\sqrt{2 D}\ \xi(t)\ ,
\label{1}
\end{equation}
with unbiased $\delta$-correlated Gaussian noise $\xi (t)$
(thermal fluctuations) of strength $D$.
The force-field $F(x,t)$ is assumed to derive from a metastable
potential with a well at $\bar x_s$ and a barrier at
$\bar x_u > \bar x_s $, subject to periodic modulations with period $\T$.
For $D=0$, the deterministic dynamics (\ref{1}) is furthermore
assumed to exhibit a stable
periodic orbit (attractor) $x_s(t)$
and an unstable periodic orbit (basin boundary) $x_u(t) > x_s(t)$.

For weak noise $D$, there is a small probability that a particle
obeying~(\ref{1}) escapes from the basin of attraction ${\cal A}(t):=(-\infty,
x_u(t)]$ of $x_s(t)$ and disappears towards infinity. For an ensemble of
particles with probability density $p(x,t)$, the population $\PA(t)$ within
the basin of attraction is $\int_{-\infty}^{x_u(t)} p(x,t)\, dx$ and the
instantaneous rate of escape $\Gamma (t)$ equals $-\dot \PA (t)/\PA (t)$.
Apart from transients at early times, this rate $\Gamma (t)$ is independent of
the initial conditions at time $t_0$. Without loss of generality we can thus
focus on $x(t_0)=x_s(t_0)$. Small $D$ implies rare escape events, i.e.\ the
deviation of $\PA (t)$ from its initial value $\PA (t_0) = 1$ is negligible.
Exploiting $\dot x_{u}(t) = F(x_{u}(t), t)$ and the Fokker-Planck-equation
$\partial_t p = \partial_x\{ - F(x,t)+D\partial_x \}\, p$ governing $p=p(x,t)$
we find for the instantaneous rate
\begin{equation}
\Gamma (t) = -D\, \partial_x p( x=x_u(t),t)\ .
\label{5}
\end{equation}
%
%
{\em Path-integral approach ---} With the choice $p(x,t_0) =
\delta(x-x_s(t_0))$ we obtain for the conditional probability $p(x,t)$ the path
integral representation \cite{path}
\begin{equation}
p (x_f, t_f) = 
\int {\cal D}x(t)\ e^{-S[x(t)]/D} \ ,
\label{6}
\end{equation}
where the ``action'' is given by 
\begin{equation}
S[x(t)] := \int_{t_0}^{t_f} dt\ [\dot x(t) - F(x(t),t)]^2 /4 
\label{7}
\end{equation}
and where $x(t_0)=x_s(t_0)$ and $x(t_f)=x_f$ are
the ``initial'' and ``final'' conditions for the paths $x(t)$.
For weak noise, the integral (\ref{6}) is dominated by a set of paths
$x^\ast_k(t)$, corresponding to minima of the action (\ref{7}) (distinguished
by the label $k$).
These satisfy an Euler-Lagrange equation 
equivalent to the following Hamiltonian dynamics
\begin{eqnarray}
\dot p^\ast_k(t) & = & -p^\ast_k(t)\, F'(x_k^\ast(t),t)\label{8}\\
\dot x^\ast_k(t) & = & 2\, p^\ast_k(t) + F(x_k^\ast(t),t)\label{9}
\end{eqnarray}
with $F'(x,t) := \partial_x F(x,t)$.
For well separated paths $x^\ast_k(t)$,
a functional saddle point approximation in (\ref{6}) yields
\begin{equation}
p (x_f, t_f)
=\sum_k \frac{\ e^{- S[x^\ast_k(t)] /D}}
{[4\pi D Q^\ast_{k}(t_f)]^{1/2}}\, [1+\Ord (D)]\ ,
\label{10}
\end{equation}
where the quantity $Q^\ast_k(t)$ relates to the determinant of fluctuations
around $x^{\ast}_{k}(t)$. Following the reasoning in \cite{schulman}, we find
for our case \cite{we} that $Q^{\ast}_{k}(t)$ obeys the relation 
\begin{eqnarray}
\ddot Q^\ast_k(t)/2 & - & 
d\, [Q^\ast_k(t)\, F'(x^\ast_k(t),t)]\, / \, dt\nonumber\\
& + &  Q^\ast_k(t)\, p^\ast_k(t)\, F''(x^\ast_k(t),t) = 0\label{11}
\end{eqnarray}
with initial conditions $Q^\ast_k(t_0)=0$ and $\dot Q^\ast_k(t_0) = 1$.
Exploiting that the derivative of the action at its end-point equals the
momentum $p^{\ast}_{k}(t_{f})$, we can infer from (\ref{5},\ref{10}) our first
main result, namely
\begin{equation}
\Gamma (t_f) = 
\sum_k \frac{p^\ast_k (t_f)\, e^{- S[x^\ast_k(t)] /D}}
{[4\,\pi\, D\, Q^\ast_{k}(t_f)]^{1/2}}\, [1+\Ord (D)]\ ,
\label{13}
\end{equation}
where the boundary conditions $x^\ast_k(t_0) = x_s(t_0)$ and
$x^\ast_k(t_f) = x_u(t_f)$ are understood in (\ref{8},\ref{9}). In
view of (\ref{10}), the instantaneous rate (\ref{13}) has the
suggestive form of probability at the separatrix times ``velocity''.

Closer inspection of (\ref{7}-\ref{9}) reveals the following
generic features of each path $x^\ast_k(t)$ which notably contributes to the 
rate (\ref{13}), see fig.~1:
Starting at $x^\ast_k(t_0) = x_s(t_0)$, it
continues to follow rather closely the stable periodic orbit $x_s(t)$ for
some time. At a certain moment, it crosses over into the vicinity of the 
unstable periodic orbit $x_u(t)$ and remains there 
for the rest of its time, ending at
$x^\ast_k(t_f) = x_u(t_f)$.  
Without loss of generality, we can sort the paths $x^{\ast}_{k}(t)$ by the
time they spend near the unstable periodic orbit, such that $x^\ast_0(t)$ is that
path which crosses over from $x_s(t)$ to $x_u(t)$ at the ``latest possible
moment''.
Apart from a time shift, each path $x^\ast_k(t)$ then closely
resembles the same ``master path'' $x^\ast(t)$ (see fig.~1). This path
$x^\ast(t)$ is defined as an absolute minimum of the action (\ref{7})
in the limit $t_0\to -\infty$, $t_f\to\infty$, and is fixed uniquely
by demanding that $x^\ast(t+k\T )$ is the ``master path'' associated with
$x^\ast_k(t)$.

The basic qualitative features of each minimizing path $x_k^\ast(t)$ are thus
quite similar to the well-known barrier-crossing problem in a static potential
\cite{caroli}. However, in the limit $t_0\to -\infty$, $t_f\to\infty$ we have,
in contrast to this latter situation, {\em not} a continuous symmetry
(Goldstone mode), but a {\em discrete degeneracy} of the minimizing paths.  
As a consequence, in our case the minimizing paths $x^\ast_k(t)$ remain 
well separated and thus the rate formula (\ref{13}) is valid for any 
(arbitrary but fixed) finite values of the driving amplitude and period, 
provided the noise strength $D$ is sufficiently small. On the other hand, for
a given $D$ we have to exclude extremely small amplitudes and extremely
long or short periods since this would lead effectively back to the static 
case.

As long as the ``master path'' $x^\ast(t)$ remains sufficiently close to the
stable periodic orbit $x_s(t)$, say for $t\leq t_s$, the force-field is well
approximated by
\begin{equation}
F(x,t)= F(x_{s}(t),t) + (x-x_{s}(t))\, F'(x_{s}(t),t)\ .
\label{15}
\end{equation}
An analogous approximation for $F(x,t)$ is valid while $x^\ast(t)$
remains in a sufficiently small neighborhood of $x_u(t)$, say for
$t\geq t_u$. The corresponding local solutions of the Hamilton equations
(\ref{8},\ref{9}) can then be written as
\begin{eqnarray}
p^\ast(t) & = & p^\ast(t_{s,u}) \, e^{-\J_{s,u}(t,t_{s,u})}\label{16}\\
x^\ast(t) & = & x_{s,u}(t)\pm p^\ast(t)\, I_{s,u}(t)\ . \label{17}
\end{eqnarray}
Here `$s,u$' means that the index is either `$s$' or `$u$' 
and the upper and lower signs in (\ref{17}) refer 
to `$s$' and `$u$', respectively.
Further, we have introduced
\begin{eqnarray}
& & \J_{s,u}(t,t_{s,u}) := \int_{t_{s,u}}^t F'(x_{s,u}(\hat t),\hat t)
\, d\hat t\label{18}\\
& & I_{s,u}(t) := \left|
2\int_{\mp\infty}^t e^{2\J_{s,u}(t,\hat t)}\, d\hat t\right| \ .\label{19}
\end{eqnarray}
Similarly, the local solutions for the prefactor in (\ref{11}) can be written as
\begin{eqnarray}
Q^\ast(t\leq t_s) & = & I_s(t)/2\label{21}\\
Q^\ast(t\geq t_u) & = & c_1/p^\ast(t)^2 - c_2\, I_u(t)\ . \label{22}
\end{eqnarray}
The parameters $p^\ast(t_{s,u})$ in (\ref{16}) and $c_{1,2}$ in (\ref{22})
cannot be fixed within such a local analysis around $x_{s,u}(t)$, they require
the global solution of (\ref{8},\ref{9},\ref{11}). We furthermore observe that 
due to the time-periodicity of $F(x,t)$ and $x_{s,u}(t)$, the quantities
\begin{equation}
\lambda_{s,u} := \J_{s,u}(t+\T,t)/\T
\label{20}
\end{equation}
are indeed $t$-independent.
The stability/instability of the periodic orbits $x_{s,u}(t)$ implies 
$\lambda_s<0$ and $\lambda_u > 0$. It follows that 
$I_{s,u}(t)$ from (\ref{19}) are  finite, 
$\T$-periodic functions.

The expressions for $x^\ast_k(t)$, $p^\ast_k(t)$, and $Q^\ast_k(t)$ 
are somewhat 
more complicated than in~(\ref{16}-\ref{22}) but since $x^\ast_k(t)$ is well 
approximated by $x^\ast(t+k\T)$, the same follows for
$p^\ast_k(t)$ and $Q^\ast_k(t)$. Closer inspection shows  \cite{we} that
in~(\ref{13}) the pre-exponential factors
$p^\ast_k(t_f)$ and $Q^\ast_k(t_f)$ can be approximated by
$p^\ast(t_f+k\T )$ and $Q^\ast(t_f+k\T)$ without further increasing the 
error $\Ord (D)$ 
in (\ref{13}). Within this same accuracy, the exponential in (\ref{13})
requires -- due to the small denominator $D$ -- a somewhat more elaborate
approximation, yielding
\begin{equation}
S[x^\ast_k(t)] = S[x^\ast(t)] +\int_{t_f+k\T}^\infty p^\ast(t)^2\, dt\ .
\label{24}
\end{equation}
{\em Rate formula ---} By introducing these approximations into (\ref{13}),
exploiting (\ref{16}-\ref{22}), and dropping the index `$f$' of $t_f$, we obtain
\cite{we} as the central result of this work the {\em instantaneous rate}
\begin{eqnarray}
& & \Gamma (t)  =  \sqrt{D}\,\alpha\, e^{-S[x^\ast(t)] / D}\,\kappa(t,D)
\, [1+E(D)]
\label{25}\\
& & \alpha  :=  [4\, \pi\,  \T^2
\lim_{t\to\infty} p^\ast (t)^2\, Q^\ast (t) ]^{-1/2} \label{26}\\
& & \kappa (t,D)  :=  \T\sum\limits_{k=-\infty}^{\infty}
\frac{\beta_k(t)^2}{D}\, e^{-\beta_k(t)^2  I_u(t)/ 2  D} \label{27}\\
& & \beta_k(t)  :=  e^{-\lambda_u k \T}
\lim_{\hat  t\to\infty} p^\ast(\hat  t )\, e^{\J_u(\hat  t,t)}\ .\label{28}
\end{eqnarray}
The relative error $E(D)$ is found to be of order
$\Ord(D)$ if $F''(x_u(t),t) \equiv 0$, and $\Ord(D^{1/2})$
otherwise.
By use of (\ref{18}-\ref{20}) one finds that the average of (\ref{27}) over a
single time-period $\T$ equals $1$.  For the {\em time-averaged rate} $\bar\Gamma$
we thus obtain
\begin{equation}
\bar\Gamma = \sqrt{D}\,\alpha\, e^{-S[x^\ast(t)]/D}\, [1+E(D)] \ .
\label{31'}
\end{equation}
It consists of an Arrhenius-type exponentially leading part and,
{\em in contrast to equilibrium rates} \cite{han90},
a non-trivial pre-exponential $D$-dependence.

{\em An archetype example ---} In general, the explicit quantitative evaluation of
$S[x^\ast(t)]$, $\alpha$, and $\kappa (t,D)$ in (\ref{25},\ref{31'}) is not
possible in closed analytical form.  An exception is the piecewise linear
force-field with additive sinusoidal driving
\begin{eqnarray}
F(x\leq 0 ,t) & = & \lambda_s\, (x-\bar x_s ) + A\,\sin(\Omega\, t)\nonumber \\
F(x\geq 0 ,t) & = & \lambda_u\, (x-\bar x_u ) + A\,\sin(\Omega\, t)\ ,
\label{2}
\end{eqnarray}
corresponding to a periodically rocked piecewise parabolic potential,
with parameters $\bar x_s < 0$, $\bar x_u > 0$, $\lambda_s < 0$,
$\lambda_u > 0$, respecting $\lambda_s \bar x_s = \lambda_u \bar x_u$
(continuity at $x=0$).
To simplify the analytical calculations we further restrict 
ourselves to the case that the ``master path'' $x^\ast(t)$ crosses the matching
point $x=0$ in (\ref{2}) only once \cite{f3}, say at $t=t_c$.
The periodic orbits $x_{s,u}(t)$ then assume the simple form
\begin{equation}
x_{s,u}(t)=\bar x_{s,u} - 
\frac{A\, [\lambda_{s,u}\,\sin(\Omega t)+\Omega \cos(\Omega t)]}
{\lambda_{s,u}^2 +\Omega^2} \ .
\label{33}
\end{equation}
Moreover, Eqs.(\ref{16},\ref{17}) are now valid with index `$s$' 
for all $t\leq t_c$ and with `$u$' for all $t\geq t_c$. 
By matching these solutions at $t=t_c$ the
global parameters $p^\ast(t_{s,u})$ in (\ref{16}) are fixed and one
obtains
\begin{equation}
t_c = \frac{1}{\Omega}\arctan
\left(\frac{\lambda_u\lambda_s - \Omega^2}
{\Omega\, (\lambda_u+\lambda_s)}\right) \ .
\label{34}
\end{equation}
To ensure that $x^\ast(t)$ is a minimum of the action in~(\ref{7})
one has
\begin{equation}
(\lambda_u + \lambda_s)\, A\, \Omega \cos(\Omega  t_c) <0\ .
\label{35}
\end{equation}
In the same manner, the prefactor (\ref{22}) has to
be matched at $t=t_c$. While $Q^\ast(t)$ is still continuous,
$\dot Q^\ast(t)$ develops a jump at $t=t_c$ which can be determined 
from (\ref{11}). Upon collecting everything, the final result reads
\begin{eqnarray}
& & S[x^\ast(t)] = \Delta V\, 
\left[ 1-\frac{|A\lambda_u\lambda_s|}{(R\, \Delta V)^{1/2}}\right]^2
\label{36}\\
& & \alpha = \left[\frac{|A|(\Omega^2+\lambda_u\lambda_s) + 
(R\, \Delta V)^{1/2}}
{16\, \pi^3\, |A|\, S[x^\ast(t)]}\right]^{1/2}\label{37}\\
& & \beta_k(t) = e^{-\lambda_u\, (k\T+t-t_c)}\,
\left[\lambda_u\bar x_u-\frac{|A\lambda_u\lambda_s|}{H^{1/2}}\right]\ ,
\label{38}
\end{eqnarray}
where $H:=(\lambda_u^2+\Omega^2)(\lambda_s^2+\Omega^2)$, 
$R:=2H/(\lambda_u^{-1}-\lambda_s^{-1})$,
and $\Delta V :=(\lambda_u\bar x_u^2 - \lambda_s\bar x_s^2)/2$
is the potential barrier corresponding to the
undriven ($A=0$) force-field (\ref{2}). 
With $\T = 2\pi /\Omega$, $I_u(t)=1/\lambda_u$, and
$t_c$ from (\ref{34},\ref{35}), the rate (\ref{25},\ref{31'}) 
is thus determined completely.

{\em Comparison ---} These analytical predictions for the
instantaneous rate (\ref{25}) are compared in fig.~2 for a
representative set of parameter values with very accurate numerical
results.  
The agreement indeed improves with decreasing noise-strength
$D$. While the absolute values of $\Gamma(t)$ and the location of the
extrema strongly depend on $D$, the overall shape changes very little
and does {\em not} develop singularities as $D\to 0$.
The corresponding time-averaged rates (\ref{31'}) are depicted in
fig.3, exhibiting excellent agreement between theory and numerics even
for relatively large $D$.  The inset of fig.~3 confirms our prediction 
that the relative error $E(D)$ in (\ref{31'}) decreases asymptotically
like $D$.
Finally, fig.4.  illustrates the dependence of the averaged rate
$\bar\Gamma$ upon the amplitude $A$ of the periodic driving force.  
As expected, our theoretical prediction compares very well with the
(numerically) exact rate, except for very small driving amplitudes
$A$ (see the discussion above eq.~(\ref{15})).  
The approximation from \cite{sme99} is complementary to ours in that
it is very accurate for small $A$ but develops considerable deviations
with increasing $A$.  
Those approximations have been omitted in figs.2,3 since they are not
valid in this parameter regime and indeed are way off.

This work was supported by
DFG-Sachbeihilfe HA1517/13-2 and
the Graduiertenkolleg GRK283.

\begin{figure}[htbp]
  \begin{center}
    \vspace*{-4.8cm} \epsfig{file=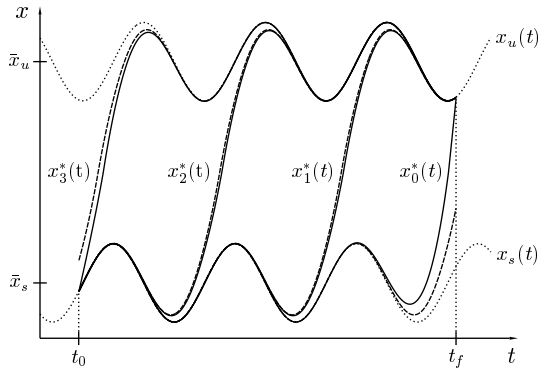,width=1.1\linewidth} \vspace*{-4.3cm}
  \end{center}
  \caption{        
    Solid: The paths $x^\ast_k(t)$, $k=0,\dots,3$, which minimize the
    action (\ref{7}) with boundary conditions $x^\ast_k(t_0)=x_s(t_0)$ and
    $x^\ast_k(t_f)=x_u(t_f)$ for the example (\ref{2}) with $\bar
    x_s=\lambda_s=-1$, $\bar x_u=\lambda_u=\Omega =1$, $A=0.5$
    (dimensionless units).  Dashed: the associated ``master paths''
    $x^\ast(t+k\,\T)$.
    Dotted: stable and unstable periodic orbits $x_s(t)$
    and $x_u(t)$.  In this plot, $t_f-t_0$ has been chosen 
    rather small. As $t_f-t_0$ increases, more and more intermediate 
    paths $x^\ast_k(t)$ appear which better and better agree with 
    $x^\ast(t+k\T)$. }
\end{figure}

\begin{figure}[htbp]
  \begin{center}
    \vspace*{-4.9cm} \epsfig{file=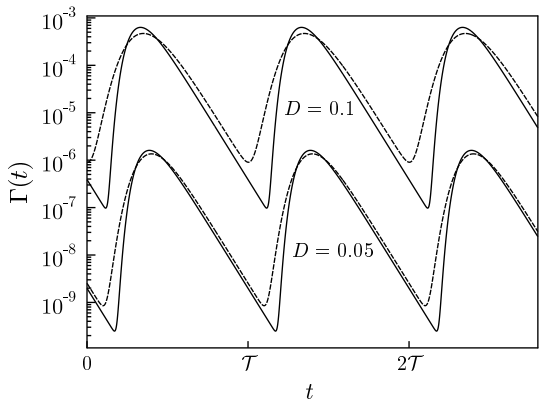,width=1.1\linewidth} \vspace*{-4.1cm}
  \end{center}
  \caption{        
    Instantaneous rate $\Gamma (t)$ for the force field (\ref{2})
    with
    $\bar x_s=\lambda_s=-1$, $\bar x_u=\lambda_u=\Omega =1$, 
    $A=0.5$, corresponding to a static ($A=0$) potential barrier $\Delta V=1$.
    Solid: analytical result (\ref{25},\ref{27},\ref{34}-\ref{38}).
    Dashed: high-precision numerical results by evolving the Fokker-Planck-equation
    for $p(x,t)$ until transients have died out and then evaluating
    (\ref{5}).}
\end{figure}

\begin{figure}[htbp]
  \begin{center}
    \vspace*{-5cm} \epsfig{file=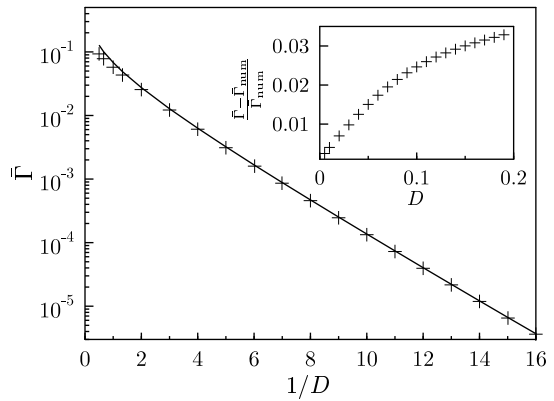,width=1.1\linewidth} \vspace*{-4cm}
  \end{center}
  \caption{        
    Arrhenius plot of the time-averaged rate $\bar\Gamma$.  
    Parameters are like in fig.2. 
    Solid: analytical result (\ref{31'},\ref{36},\ref{37}).
    Crosses: precise numerical results. 
    Inset: relative difference between
    analytical ($\bar\Gamma$) and numerical ($\bar\Gamma_{\rm num}$)
    rate. }
\end{figure}

\begin{figure}[htbp]
  \begin{center}
    \vspace*{-4.9cm} \epsfig{file=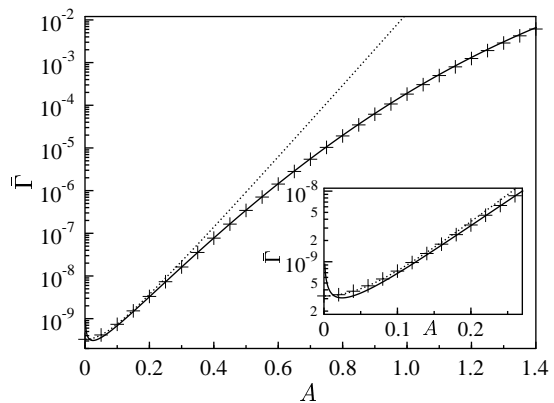,width=1.1\linewidth} \vspace*{-4.1cm}
  \end{center}
  \caption{        
    Time-averaged rate $\bar\Gamma$ vs. driving amplitude $A$ for $D=0.05$.
    Parameters are like in fig.2.
    Crosses: precise numerical results. 
    Solid: analytical result (\ref{31'},\ref{36},\ref{37}).
    Dotted: theoretical approximation from Ref.~[4]. 
    The small-$A$ regime is magnified in the inset.}
\end{figure}


\begin{thebibliography}{99}
\bibitem{f3} In general, the self-consistency of the final solution $x^\ast(t)$ 
with this assumption has to be checked numerically (a transcendental equation
arises). Analytically, one can show that $A^2 < \lambda_{s,u}^2\,\bar x_{s,u}^2$ 
is a sufficient (but not necessary) condition, while
$A^2 < (\lambda_{s,u}^2+\Omega^2)\,\bar x_{s,u}^2$ is necessary but not 
sufficient.
The latter inequality also guarantees that the
expressions in the square brackets in (\ref{36}-\ref{38}) 
are positive.
\end{thebibliography}
\end{document}